\documentclass{article}
\usepackage{dcolumn}% Align table columns on decimal point
\usepackage{bm}% bold math
\usepackage{verbatim}       % Defines \begin{comment}, \end{comment}

\usepackage[dvips]{graphicx}
\usepackage{amsthm}
\usepackage{amssymb}
\usepackage{bm}
\usepackage{amstext}
\usepackage{psfrag}
\usepackage{amsmath}
\usepackage{amsfonts}

\newfont{\bb}{msbm10 at 12pt}

\newcommand{\dd}{{\rm d}}
\newcommand{\bd}{\begin{definition}}                %inizia definizione
\newcommand{\ed}{\end{definition}}                  %fine definizione
\newcommand{\bc}{\begin{corollary}}                 %inizia corollario
\newcommand{\ec}{\end{corollary}}                   %fine corollario
\newcommand{\bl}{\begin{lemma}}                     %inizia lemma
\newcommand{\el}{\end{lemma}}                       %fine lemma
\newcommand{\bp}{\begin{proposition}}            %inizia proposizione
\newcommand{\ep}{\end{proposition}}                %fine proposizione
\newcommand{\bere}{\begin{remark}}                  %inizia osservazione
\newcommand{\ere}{\end{remark}}                     %fine oservazione

\newcommand{\bt}{\begin{theorem}}
\newcommand{\et}{\end{theorem}}

\newcommand{\be}{\begin{equation}}
\newcommand{\ee}{\end{equation}}

\newcommand{\bit}{\begin{itemize}}
\newcommand{\eit}{\end{itemize}}
\newtheorem{theorem}{Theorem}[section]
\newtheorem{corollary}[theorem]{Corollary}

\newtheorem{lemma}[theorem]{Lemma}
\newtheorem{proposition}[theorem]{Proposition}
\theoremstyle{definition}
\newtheorem{definition}[theorem]{Definition}
\theoremstyle{remark}
\newtheorem{remark}[theorem]{Remark}

\begin{document}
%
%\DeclareGraphicsExtensions{.pdf}

%\title{On the relationship between $K$-causality and infinite $A$-causality}
\title{The causal ladder and the strength of $K$-causality. I}

\author{E. Minguzzi \footnote{Department of Applied Mathematics, Florence
 University, Via S. Marta 3,  50139 Florence, Italy. E-mail: ettore.minguzzi@unifi.it}}

\date{}
\maketitle

\begin{abstract}
\noindent A unifying framework for the study of causal relations is
presented. The causal relations are regarded as subsets of $M\times
M$ and the role of the corresponding antisymmetry conditions in the
construction of the causal ladder is stressed. The causal hierarchy
of spacetime is built from chronology up to $K$-causality and new
characterizations of the distinction and strong causality properties
are obtained. The closure of the causal future is not transitive, as
a consequence its repeated composition leads to an infinite causal
subladder between strong causality and $K$-causality - the
$A$-causality subladder. A spacetime example is given which proves
that $K$-causality differs from infinite $A$-causality.

\end{abstract}

%\pacs{04.20.-q, 04.20.Gz, 04.20.Cv}
%MSC: 83C75

%\noindent Key Words:

\section{Introduction}
The causal relations are usually presented through their point based
counterparts, namely the sets $I^{\pm}(x)$, $J^{\pm}(x)$, however
the most natural and effective approach regards them as subsets of
$M\times M$. It is convenient to define \cite{minguzzi06c} the
following sets on $M \times M$
\[
I^{+}=\{(p,q) : p\ll q  \} , \quad J^{+}=\{(p,q) : p\leq q  \},
\quad E^{+}=\{(p,q) : p\rightarrow q  \}.
\]
Clearly, $E^+={J^+}\backslash I^+$. Moreover, $I^{+}$ is open
\cite[Chap. 14, Lemma 3]{oneill83} \cite[Prop. 2.16]{minguzzi06c},
$\bar{J}^+=\bar{I}^+$, $\textrm{Int}\, J^+=I^+$ and
$\dot{J}^+=\dot{I}^+$ \cite[Prop. 2.17]{minguzzi06c}. Once these
sets are defined the conditions of chronology or causality are
obtained as antisymmetry conditions on the corresponding relations
on $M\times M$.

The approach with sets on $M\times M$ is also useful for the
definition of new causal relations. For instance, about ten years
ago Sorkin and Woolgar \cite{sorkin96} defined the relations $K^{+}$
as the smallest closed subset $K^{+}\subset M\times M$, which
contains $I^{+}$, $I^{+}\subset K^{+}$, and shares the transitivity
property: $(x,y)\in K^{+}$ and $(y,z) \in K^{+}$ $\Rightarrow (x,z)
\in K^{+}$ (the set of  causal relations satisfying these properties
is non-empty, consider for instance the trivial subset $M\times M$).
This definition raised from the fact that $J^{+}$ while transitive
is not necessarily closed whereas $\bar{J}^{+}$ while closed is not
necessarily transitive.

We shall see other examples in this work where the approach on
$M\times M$ has proved not only useful but also superior to the one
with the point based relations. For instance, the $A$-causality
subladder introduced by Penrose will prove more natural than
Carter's causal virtuosity subladder.

The aim of this work is to present a unifying framework for all the
causal relations that have appeared in the literature. The various
causality conditions are then traced back to conditions (usually to
antisymmetry conditions) on these causal relations  and the
relationship between  the different causality requirements becomes
trivial and related to the inclusion of sets on $M\times M$.

Actually, since stable causality can be regarded as an antisymmetry
condition on the Seifert future $J^{+}_S$, it could also be included
in the present study. However, since there is an open issue as to
whether stable causality coincides with $K$-causality I have
preferred to leave these questions to a related work where the
mentioned problem is studied in deep \cite{minguzzi07}.

The work is organized as follows.

In section \ref{mja} a general approach to causal relations as
subset of $M\times M$ initiated in \cite{minguzzi06c} is introduced.
The role played by the antisymmetry condition is stressed in view of
its unifying role for the construction of the causal ladder. Not all
the definitions or results presented in this section are later used.
They are given because they hold whatever the causal relation
considered and because the section is intended as a reference for
future work, for instance for \cite{minguzzi07}.

In section \ref{nbe} new characterizations of the distinction and
strong causality properties are obtained. In particular the past
(resp. future) distinction is proved to follow from the antisymmetry
of a causal relation termed $D_p$ (resp. $D_f$). Strong causality is
not characterized through the antisymmetry of a causal relation but,
nevertheless, a similar useful result is obtained (theorem
\ref{mka}).

In section \ref{cxs} the causal relations coming from the successive
composition of $\bar{J}^{+}$ are considered. They give rise to a
causal subladder which I  clarify mentioning the different
definitions that can be found in the literature. I shall mainly use
the approach due to Penrose \cite[Remark 4.19]{penrose72}, who
described the ladder explicitly, though  I will not use the same
terminology.

\begin{figure}[ht]
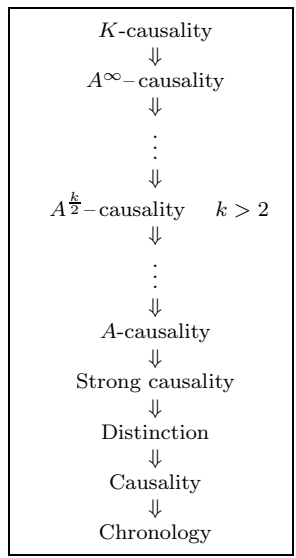

\begin{center} \fbox{ {\footnotesize
\begin{tabular}{c}
\vspace{-0.3cm}\\
{$K$-causality} \\
$\Downarrow $ \\
{$A^{\infty}$--$\,$causality} \\
$\Downarrow $ \\
$\vdots $ \\
$\Downarrow $ \\
\ {$A^{\frac{k}{2}}$--$\,$causality} \quad $k>2$ \\
$\Downarrow $ \\
$\vdots $ \\
$\Downarrow $ \\
{$A$-causality} \\
$\Downarrow $ \\
{Strong causality} \\
$\Downarrow $ \\
{Distinction} \\
$\Downarrow $ \\
{Causality} \\
$\Downarrow $ \\
{Chronology}\\
\end{tabular}}}
\end{center}
\caption{A portion of the causal ladder of spacetimes. An arrow
between two properties $\mathcal{P}_1 \Rightarrow \mathcal{P}_2$
means that the former implies the latter and that there are examples
of spacetimes in which the latter holds and the former does not
hold.
%Carter's
%ladder of $n$-th degree ($n\ge3$) causal virtuosity is not displayed
%and would stay between strong causality and $K$-causality.
}
\label{ladder}
\end{figure}

In   section \ref{bap}  I provide an example of spacetime which is
infinite $A$-causal and yet non-$K$-causal. This result was long
suspected but it had remained open so far. The information provided
by the portion of the causal ladder displayed in figure \ref{ladder}
is then accurate and complete.

I refer the reader to \cite{minguzzi06c} for most of the conventions
used in this work. In particular, I denote with $(M,g)$ a $C^{r}$
spacetime (connected, time-oriented Lorentzian manifold), $r\in \{3,
\dots, \infty\}$ of arbitrary dimension $n\geq 2$ and signature
$(-,+,\dots,+)$. On $M\times M$ the usual product topology is
defined.

%I use the short-hand notation $\bar{J}^{+}(x)=\overline{J^{+}(x)}$.

The subset symbol $\subset$ is reflexive, $X \subset X$.  With
$J^{+}_U\subset U\times U$, I denote the causal relation on the
spacetime $U$ with the induced metric, so that $x\le_U z$ reads
$(x,z) \in J^{+}_U$.  By {\em neighborhood} it is always meant an
open set. The boundary of a set $B$ is denoted $\dot{B}$.

\section{Preliminaries} \label{mja}

Due to the proliferation of causal relationships it is desirable to
provide a common framework since many results can be derived in
similar ways.  The definitions given here are for most part
compatible with those used in set theory \cite{hrbacek99}. A
(binary) {\em relation} is a subset $R^{+} \subset M\times M$, it is
a {\em causal relation} if it contains $I^{+}$, $I^{+}\subset
R^{+}$, where $I^{+}$ is the chronological future for some given
Lorentzian metric on $M$. It is {\em open} (resp. {\em closed}) if
$R^{+}$ is open (resp. closed) as a subset of  $M \times M$.  Given
$R^{+}_1$ and $R^{+}_2$ the composition $R^{+}_{2}\circ R^{+}_1$ is
the set
\[
R^{+}_{2}\circ R^{+}_1=\{(x,z) \in M\times M:  \exists  y, (x,y) \in
R^{+}_1 \textrm{ and } (y,z) \in R^{+}_2 \}
\]
The diagonal  $\Delta=\{(x,x), x \in M\}$ is an identity for this
composition that is, whatever $R^{+}$, $\Delta \circ
R^{+}=R^{+}\circ \Delta=R^{+}$.
 The relation $R^{+}$ is {\em transitive} if
$R^{+} \circ R^{+} \subset R^{+}$, that is for all $x,y,z \in M$,
$(x,y) \in R^{+}$ and $(y,z) \in R^{+}$ $\Rightarrow (x,z) \in
R^{+}$. It is idempotent if $R^{+} \circ R^{+} = R^{+}$ (for
instance $I^{+}$ and $J^{+}$ are idempotent). It is {\em reflexive}
if for all $x \in M$, $(x,x) \in R^{+}$, that is $\Delta \subset
R^{+}$. It is {\em irreflexive} if for all $x \in M$, $(x,x) \notin
R^{+}$, that is $R^{+}\cap \Delta=\emptyset$. Clearly, every
reflexive and transitive relation is idempotent. Note also that
$\circ$ has a distributive property with respect to arbitrary unions
of sets $\bigcup _\alpha A_{\alpha}^+ \circ \bigcup _\beta
B_{\beta}^+=\bigcup _{\alpha,\beta} ( A_{\alpha}^{+} \circ
B_{\beta}^+) $. The properties of reflexivity and transitivity are
also invariant under arbitrary intersections of relations sharing
them.

 A set $B^{+}$ is a left (resp.
right) $R^{+}$-{\em ideal} if $B^{+} \subset R^{+}$ and $R^{+} \circ
B^{+} \subset B^{+}$ (resp. $B^{+} \circ R^{+} \subset B^{+}$), and
an $R^{+}$-ideal if it is both a left and a right $R^{+}$-ideal (for
instance $I^{+}$ is a $J^{+}$-ideal).

The relation $R^{+}$  is {\em antisymmetric} if for all $x,z \in M$,
\[ (x,z) \in R^{+} \textrm{ and } (z,x) \in R^{+} \Rightarrow x=z\]
in which case $M$ is said to be $R${\em -causal} ($I$-causality
coincides with chronology\footnote{Indeed, if chronology holds then
the antisymmetry condition holds because the hypothesis ``$ (x,z)
\in I^{+}$ and $(z,x) \in I^{+}$'' is false for every $x,z \in M$.
Conversely, if the antisymmetry condition holds then no closed
timelike curve can exist otherwise one could find $x\ne z$, such
that the hypothesis of the antisymmetry condition holds.} and
$J$-causality coincides with causality). {\em $R$-causality at} $x
\in M$ holds, if there is no point $z\in M$, $z \ne x$ such that $
(x,z) \in R^{+}$ and $(z,x) \in R^{+}$. An important observation is
that if $A^{+} \subset B^{+}$ and $B^{+}$ is antisymmetric then
$A^{+}$ is antisymmetric too. $R^{+}$ is {\em asymmetric} if $
\forall x,z \in M$, $(x,z) \in R^{+} \Rightarrow (z,x) \notin
R^{+}$. Thus asymmetry is equivalent to antisymmetry and
irreflexivity. Transitivity and irreflexivity imply asymmetry.

$R^{+}$ is a non-strict (or reflexive) {\em partial order} if it is
reflexive, transitive and antisymmetric. It is a strict  (or
irreflexive) partial order if it is irreflexive and transitive.
There is a one-to-one correspondence between non-strict and strict
partial orders obtained by including or removing the diagonal.
Indeed, the reader may easily prove that given the binary relation
$R^{+}$, the reflexive relation $R^{+}\cup \Delta$ is transitive and
antisymmetric iff the irreflexive relation $R^{+}\cap \Delta^{C}$ is
transitive. Finally, strict partial orders can be characterized also
as those relations which are asymmetric and transitive.

The reader may check that the next definition coincides with the
usual  one\cite{kronheimer67}
\begin{definition}
A triple $(M,R^{+},B^{+})$, where $M$ is a set, $R^{+}\subset
M\times M$ is a reflexive partial order and $B^{+}\subset R^{+}$ is
a irreflexive $R^{+}$-ideal is a {\em causal structure in the sense
of Kronheimer and Penrose}.
\end{definition}

Given $R^{+}$, the relation $R^{-}\subset M\times M$ is given by the
set
\begin{equation}
R^{-}=\{(x,z) \in M\times M: (z,x)\in R^{+}\},
\end{equation}
in particular in the notation of the ``$R$-causality'' property the
sign is omitted because the antisymmetric condition for $R^{+}$
coincides with that for $R^{-}$. Analogously, the diagonal $\Delta$
does not exhibit the plus sign because $\Delta^{-}=\Delta$.

Given $R^{+}$ the point based relations $R^{+}(x)$ and $R^{-}(x)$
are defined as

\begin{align}
R^{+}(x)&=\{y \in M: (x,y) \in R^{+}\}, \\
R^{-}(x)&=\{y \in M: (y,x) \in R^{+}\}.
\end{align}
Note that $y\in R^{+}(x)$ iff $(x,y)\in R^{+}$ iff $x \in R^{-}(y)$.

$R^{+}$ is {\em partially open} (resp. {\em partially closed}) if
for all $x \in M$ the sets $R^{+}(x)$ and $R^{-}(x)$ are open (resp.
closed). Note that provided $I^{+} \subset R^{+}$, if $R^{+}$ is
partially closed then it is reflexive, because $x \in
\overline{I^{+}(x)}\subset R^{+}(x)$. It is trivial to prove that if
$R^{+}$ is open (resp. closed) then it is partially open (resp.
partially closed). Remarkably, the converse also holds provided
$I^{+}\subset R^{+}$ and $R^{+}$ is transitive as then next result
shows.

\begin{theorem} \label{cqo}
Let $(M,g)$ be a spacetime and let $I^{+}$ be the chronological
relation. A transitive causal ($I^{+}\subset R^{+}$) relation  is
open (resp. closed) iff it is partially open (resp. partially
closed). Moreover, in this case it is also idempotent.
\end{theorem}

\begin{proof}
Assume $R^{+}$ partially open (resp. partially closed) the other
direction
being trivial.\\
Open set case. Assume that $R^{+}$ is partially open and let $(x,z)
\in R^{+}$ so that $x \in R^{-}(z)$. First let me show that $R^{+}$
is idempotent. Indeed, since it is partially open there is $y \in
I^{+}(x) \cap R^{-}(z)$, and since $I^{+}\subset R^{+}$, it is
$I^{+}(x) \subset R^{+}(x)$, and finally $(x,y) \in R^{+}$ and
$(y,z)\in R^{+}$, which proves $R^{+}\circ R^{+}=R^{+}$. Now,
consider again arbitrary $(x,z) \in R^{+}$. Since $R^{+}$ is
partially open there is a open neighborhood $U\ni x$ of compact
closure such that $\bar{U} \subset R^{-}(z)$. Since $R^{+}$ is
idempotent, $\bar{U}$ is covered by the open sets $\{R^{-}(y), \quad
y\in R^{-}(z)\}$. Thus there  is a finite number of events $y_i \in
R^{-}(z)$, $i=1,\ldots, n$, and a subcovering of $\bar{U}$,
$\{R^{-}(y_1), \ldots, R^{-}(y_n)\}$. Define the open set
$V=\bigcap_i R^{+}(y_i)$ so that $z \in V$, then for every $\bar{x}
\in U$ and $\bar{z} \in V$, $(\bar{x},\bar{z}) \in R^{+}$.

Closed set case. As already mentioned partial closure together with
$I^{+}\subset R^{+}$ implies reflexivity which implies the
idempotent property. Let $(x,z)\in \bar{R}^{+}$ and let $(x_n,z_n)
\in R^{+}$, $(x_n,z_n) \to (x,z)$. Take $s \in I^{-}(x)$ so that $x
\in I^{+}(s)$. For large $n$, $x_n \in I^{+}(s)$ and hence $x_n \in
R^{+}(s)$. By transitivity $z_n \in R^{+}(s)$ and by partial closure
$z \in R^{+}(s)$, thus $s \in R^{-}(z)$ and taking the limit $s \to
x$, again by partial closure, $(x,z)\in R^{+}$.
\end{proof}

With $R^{+}(x,y)$ it is  denoted the set $R^{+}(x) \cap R^{-}(y)$. A
set $V\subset M$ is $R${\em -convex} if the causal relation $R^{+}$
is such that for every $x,z\in V$, $R^{+}(x,y) \subset V$
($J$-convexity is the usual causal convexity). A spacetime  is {\em
strongly $R$-causal at $x\in M$} if $x$ admits arbitrarily small
$R$-convex neighborhoods, that is, for every open set $U\ni x$ there
is a $R$-convex neighborhood $V\subset U$, $x \in V$. The spacetime
is {\em strongly $R$-causal} if it is strongly $R$-causal at every
point (strong $J$-causality coincides with strong causality).

$R^{+}$ is {\em injective} if the maps on the set of parts of $M$,
$R^{\pm}:M \to P(M)$, are injective, that is, $R^{+}(x)=R^{+}(z)
\Rightarrow x=z$, and analogously for $R^{-}$.

\begin{theorem} \label{cfw} The generic relation $R^{+}$ satisfies
\begin{itemize}
\item[(a)]  If  $R^{+}:M \to P(M)$ or $R^{-}:M \to P(M)$  are injective and $R^{+}$ is transitive then $R^{+}$ is antisymmetric.
\item[(b)] If $R^{+}$ is antisymmetric and reflexive then both maps $R^{+}:M \to P(M)$ and $R^{-}:M \to P(M)$  are  injective.
\item[(c)] If $R^{+}$ is transitive and reflexive then the
injectivity of the map $R^{+}:M \to P(M)$ is equivalent to the
injectivity of the  map $R^{-}:M \to P(M)$. Moreover, the
injectivity is equivalent to the antisymmetry.
%\item[(d)] If $R^{+}$ is partially open or partially closed then the
%injectivity of the map $x \to R^{+}(x)$ is equivalent to the
%injectivity of the map $x \to \bar{R}^{+}(x)$, which is equivalent
%to the injectivity of the map $x \to \textrm{Int}\,R^{+}(x)$ (an
%analogous statement holds for the minus case).
\end{itemize}
\end{theorem}

\begin{proof}
Proof of (a). Assume $x \to R^{+}(x)$ is injective and that $R^{+}$
is transitive. Take $x,z \in M$ such that $(x,z)\in R^{+}$ and
$(z,x) \in R^{+}$. Let $y \in R^{+}(x)$, since $R^{+}$ is transitive
$y\in R^{+}(z)$, thus $R^{+}(x) \subset R^{+}(z)$. The other
inclusion is analogous thus $R^{+}(x)=R^{+}(z)$ and by injectivity
$x=z$.

Proof of (b). Assume $R^{+}$ is antisymmetric and reflexive and take
$x,z \in M$ such that $R^{+}(x)=R^{+}(z)$. Then, because of
reflexivity $x \in R^{+}(x)=R^{+}(z)$ and analogously $z \in
R^{+}(x)$, thus by antisymmetry $x=z$.

Proof of (c). It is a trivial consequence of (a) and (b).

%Proof of (d). Trivial, if $R^{+}$ is partially open
%$R^{+}(x)=R^{+}(y)$ iff $\bar{R}^{+}(x)=\bar{R}^{+}(y)$ (and
%analogously in the partially closed case).
\end{proof}
This theorem shows that under the assumption of transitivity and
reflexivity the injectivity is equivalent to the antisymmetry and
hence $R$-causality can be expressed in terms of the injectivity of
the point based maps $R^{+}(x)$. That most of the causality
conditions can be restated as an injectivity condition on some
suitable causality maps from $M$ to $P(M)$ has  been checked in
detail by I. R\'acz \cite{racz87}. The same happens for
$K$-causality because, since $K^{+}$ is closed (and hence reflexive)
and transitive by definition, it follows from theorem \ref{cfw} that
\begin{corollary} \label{co1}
The spacetime $(M,g)$ is $K$-causal iff the map $x \to K^{+}(x)$ (or
$x \to K^{-}(x)$) is injective.
\end{corollary}
Theorem \ref{cqo} implies

\begin{corollary} \label{co2}
$K^{+}$ is the smallest transitive relation containing $I^{+}$ such
that for every $x \in M$, $K^{+}(x)$ and $K^{-}(x)$ are closed.
\end{corollary}

\begin{proof}
It follows from the fact that $K^{+}$ is the intersection of all the
relations which are transitive, contain $I^{+}$ and are closed which
by theorem \ref{cqo} is the intersection of all the relations which
are transitive, contain $I^{+}$ and are partially closed.
\end{proof}

\section{Distinction and strong causality} \label{nbe}
In this section new characterizations of the distinction and strong
causality properties are obtained.

A spacetime is future (resp. past) {\em distinguishing} if
$I^{+}(x)=I^{+}(z)$ (resp. $I^{-}(x)=I^{-}(z)$) $\Rightarrow x=z$.
For other characterizations not considered here see \cite[Lemma
3.10]{minguzzi06c}).

\begin{theorem} \label{pxs}
The spacetime $(M,g)$ is future (resp. past) distinguishing if and
only if for every $x,z \in M$, $(x,z) \in {J}^{+}$ and $x \in
\overline{J^{+}(z)}$ imply $x=z$ (resp.  $(x,z) \in {J}^{+}$ and $z
\in \overline{J^{-}(x)}$ imply $x=z$).
\end{theorem}
\begin{proof}
(Future case, the past case being analogous). If there is $x\ne z$
such that $(x,z) \in {J}^{+}$ and $x \in \overline{J^{+}(z)}$ then
because of $(x,z) \in {J}^{+}$, $I^{+}(z) \subset I^{+}(x)$ while
because of $x \in \overline{J^{+}(z)}$, $I^{+}(x) \subset I^{+}(z)$,
thus $I^{+}(x) = I^{+}(z)$, that is $(M,g)$ is not future
distinguishing.

Conversely, if $(M,g)$ is not future distinguishing there is $x' \ne
z$ such that $I^{+}(x')=I^{+}(z)$. Since $z \in
\overline{I^{+}(z)}=\overline{J^{+}(x')}$, let $\sigma_n$ be a
sequence of causal curves of endpoints $x'$ and $z_n$, $z_n \to z$,
and let $\sigma^{z}$ be a limit curve of the sequence passing
through $z$. Take $x \in \sigma^{z}\backslash \{z\}$, then $(x,z)
\in J^{+}$ and $x \in \overline{J^{+}(x')}=\overline{J^{+}(z)}$, but
$x \ne z$.

\end{proof}

It is convenient to introduce the following subsets of $M \times M$,
\begin{align}
D^{+}_f&=\{(x,y): y \in \overline{I^{+}(x)}\,\},\\
D^{+}_p&=\{(x,y): x \in \overline{I^{-}(y)}\,\}.
\end{align}
Clearly, $J^{+}\subset D_f^{+} (\textrm{or } D_p^{+}) \subset
\bar{J}^{+}$.

\begin{definition}
A spacetime is future (resp. past) reflecting   if
$D^{+}_f=\bar{J}^{+}$ (resp. $D^{+}_p=\bar{J}^+$). Equivalently, the
spacetime is future (resp. past) reflecting if $(x,z) \in
\bar{J}^{+} \Leftrightarrow z \in \bar{J}^{+}(x)$ (resp. $(x,z) \in
\bar{J}^{+} \Leftrightarrow x \in \overline{J^{-}(z)}$). A spacetime
is reflecting if it is both past and future reflecting.
\end{definition}

The equivalence with other more traditional definitions of
reflectivity follows from \cite[Prop. 3.45]{minguzzi06c} and
\cite[Prop. 1.3]{hawking74}). Note the different meanings of  the
terms {\em reflecting} which refers to the spacetime, and {\em
reflexive} which refers to the causal relations.

\begin{theorem} \label{mka}
The causal relations $D^{+}_f$ and $D^{+}_p$ are reflexive and
transitive. Moreover, $D^{+}_f$ (resp. $D^{+}_p$) is antisymmetric,
and hence a partial order, iff the spacetime is future (resp. past)
distinguishing.
\end{theorem}

\begin{proof}
The reflexivity is trivial because $x \in \overline{I^{\pm}(x)}$.
The transitivity has been proved in \cite[Claim 1]{dowker00}. In the
future case the proof  goes as follows, if $y \in
\overline{I^{+}(x)}$ and $z \in \overline{I^{+}(y)}$, taken $w \in
I^{+}(z)$, $z \in I^{-}(w)$ and since $I^{+}$ is open $y\in I^{-}(w)
$, and again since $I^{+}$ is open $x \in I^{-}(w)$, $w \in
I^{+}(x)$. Since $w$ can be chosen arbitrarily close to $z$, $z \in
\overline{I^{+}(x)}$. The antisymmetry of $D^{+}_f$ is equivalent to
the injectivity of the map $x \to \overline{I^{+}(x)}$ because of
theorem \ref{cfw} point (c). Finally, since
$\overline{I^{+}(x)}=\overline{I^{+}(y)}$ $\Leftrightarrow
{I}^{+}(x)={I}^{+}(y)$, the said injectivity is equivalent the
future distinction of the spacetime.
\end{proof}

Stated in another way, future distinction is equivalent to
$D_f$-causality and past distinction is equivalent to
$D_p$-causality. Note that this result is in one direction weaker
than theorem \ref{pxs} while in the other it is stronger. It implies
that if there is a pair of distinct events such that $z \in
\overline{I^{+}(x)}$ and $x \in \overline{I^{+}(z)}$ then there is
another such that $(x',z') \in {J}^{+}$ and $x' \in
\overline{I^{+}(z')}$.

A spacetime $(M,g)$ is {\em strongly causal} at $x$ if it admits
arbitrarily small causally convex neighborhoods at $x$. It is
strongly causal if it is strongly causal at every event.

\begin{theorem} \label{cvd}
The spacetime $(M,g)$ is strongly causal if and only if  for every
$x,z \in M$, $(x,z) \in {J}^{+}$ and $(z,x) \in \bar{J}^{+}$ imply
$x=z$.

In particular, if $(x,z) \in {J}^{+}$, $(z,x) \in \bar{J}^{+}$ and
$x \ne z$ then at the events belonging to $J^{+}(x)\cap J^{-}(z)$
the spacetime is non-strongly causal.
\end{theorem}

\begin{proof}
Assume that $x \ne z$ but  $(x,z) \in {J}^{+}$ and $(z,x) \in
\bar{J}^{+}$, and let us prove that $(M,g)$ is not strongly causal
at $r \in J^{+}(x)\cap J^{-}(z)$. Let $U\ni r$ be an arbitrary small
neighborhood whose closure does not contain both $x$ and $z$. Take
$y \in I^{+}(r)\cap U$ and $w \in I^{-}(r)\cap U$, then $z \in
I^{+}(w)$ and $x \in I^{-}(y)$. If $\sigma_n$ is a sequence of
causal curves of endpoints converging respectively to $z$ and $x$
then for sufficiently large $n$ the first endpoint stays in
$I^{+}(w)$ while the second endpoint stays in $I^{-}(y)$. As a
result $y \in I^{+}(w)$  and a timelike curve connecting $y$ to $w$
can be chosen that passes arbitrary close to $z$ and $x$ and hence
is not entirely contained in $U$. Thus the spacetime is not strongly
causal at $r$.

Conversely, if $(M,g)$ is not strongly causal then the
characterizing property (ii) of \cite[Lemma 3.22]{minguzzi06c} does
not hold, that is, there is $x \in M$, a neighborhood $U \ni x $ and
a sequence of causal curves $\sigma_n$, not entirely contained in
$U$, of endpoints $x_n,z_n$, with $x_n\to x$, $z_n \to x$. Let $C\ni
x$ be a convex neighborhood whose compact closure is contained in
another convex neighborhood $V \subset U$. Let $c_n\in \dot{C}$ be
the first point at which $\sigma_n$ escapes $C$. Since $\dot{C}$ is
compact there is $c \in \dot C$, and a subsequence such that $c_k
\to c$ and since $V$ is convex, the causal relation on $V\times V$,
$J^{+}_V$, is closed and hence $(x,c) \in J^{+}_V$ thus $(x,c) \in
J^{+}$. But since $(c_k,z_k) \in J^{+}$ it is $(c,x) \in
\bar{J}^{+}$ and yet $c \ne x$.
\end{proof}

A related result is \cite[Theor. 4.31]{penrose72}. A trivial
consequence of theorems \ref{pxs} and \ref{cvd} is
\begin{corollary}
If $(M,g)$ is strongly causal then it is distinguishing.
\end{corollary}

A consequence of $\bar{J}^{+}\subset K^{+}$ and theorem \ref{cvd} is

\begin{corollary}
If $(M,g)$ is $K$-causal then it is strongly causal.
\end{corollary}

Actually, a stronger result holds (theorem \ref{nj}).

\begin{theorem} \label{psf}
If a spacetime $(M,g)$ is future reflecting (resp. past reflecting)
 then $\bar{J}^{+}=K^{+}=D^{+}_f$ (resp.
$\bar{J}^{+}=K^{+}=D_p^{+}$). Moreover, if it is also future
distinguishing (resp. past distinguishing) then it is $K$-causal and
thus distinguishing.
\end{theorem}

\begin{proof}
I give the proof in the future case. Future reflectivity reads
$D^{+}_f=\bar{J}^{+}$ thus $\bar{J}^{+}$ is not only closed but also
transitive, and it is the smallest relations with these properties
containing $I^{+}$ hence $\bar{J}^{+}=K^{+}=D^{+}_f$. If the
spacetime is also future distinguishing then $D^+_f=K^{+}$  is
antisymmetric, i.e. the spacetime is $K$-causal.
\end{proof}

\section{The $A$-causality subladder} \label{cxs}
The set $\bar{J}^{+}\subset M\times M$ defines a causal relation
which, following Woodhouse \cite{woodhouse73}, it is also denoted
$A^{+}$. Consistently with Woodhouse's notations and in agreement
with the general definitions of section \ref{mja}, I define the {\em
almost causal} future and past of an event $x\in M$ as follows
\begin{align}
A^{+}(x)&=\{y \in M: (x,y) \in \bar{J}^{+}\}, \\
A^{-}(x)&=\{y \in M: (y,x) \in \bar{J}^{+}\}.
\end{align}
These sets are clearly closed on $M$. According to
Woodhouse\footnote{As a matter of fact Woodhouse did not use this
terminology. Note that the terminology for $A^{+}=\bar{J}^{+}$ and
$A^{\pm}(x)$, which are called {\em almost causal} futures, suggests
to call the property of $A$-causality as  {\em almost causality}.
This terminology would follow by analogy with the causal and
chronology conditions which can be expressed as antisymmetric
conditions on the chronological and causal futures. Unfortunately,
this terminology would suggest that $A$-causality is a weaker
condition than causality while it is actually stronger. Thus I keep
only the term $A${\em -causality} which is also consistent with the
general definitions at the beginning of section \ref{mja}. Note that
J.C. Park \cite{park97} called {\em almost causal} a spacetime which
satisfies the relation $\bar{J}^{+}=J^{+}_S$. This terminology is
not used here. } a spacetime is $A$-causal (or $W$-causal) if the
causal relation $A^{+}=\bar{J}^{+}$ on $M \times M$ is
antisymmetric, that is if $(x,y)\in \bar{J}^{+}$ and $(y,x) \in
\bar{J}^{+}$ imply $x=y$. Clearly, since $J^{+}\subset \bar{J}^{+}
\subset K^{+}$ (recall theorem \ref{cvd})

\begin{theorem} \label{nj}
If $(M,g)$ is $K$-causal then it is $A$-causal. Moreover, if $(M,g)$
is $A$-causal then it is strongly causal.
\end{theorem}

That the converse of these statements does not hold is shown by a
classical example due to Carter \cite[Fig. 25]{penrose72}.

The {\em common } future $\uparrow S$ and past $\downarrow S$ of a
set $S\subset M$ are open sets defined as follows
\begin{align}
\uparrow S&=\textrm{Int}\{\bigcap_{x\in S}I^{+}(x) \} =\textrm{Int}\{z\in M: \forall s \in S, s \ll z \}, \\
\downarrow S&=\textrm{Int}\{\bigcap_{x\in S}I^{-}(x) \}
=\textrm{Int}\{z\in M: \forall s \in S, z \ll s \}.
\end{align}
Note that $I^{+}(x) \subset \uparrow I^{-}(x)$ and $I^{-}\subset
\downarrow I^{+}(x)$. It is not difficult to prove \cite[Prop.
3]{akolia81} that $A^{+}(x)=\overline{\uparrow I^{-}(x)}$ and
$A^{-}(x)=\overline{\downarrow I^{+}(x)}$, and hence, since
$\uparrow I^{-}(x)$ and $\downarrow I^{+}(x)$ are open by
definition, $\uparrow I^{-}(x)=\textrm{Int} \,A^{+}(x)$ and
$\downarrow I^{+}(x)=\textrm{Int} \,A^{-}(x)$. Actually, a stronger
result holds

\begin{lemma} \label{hyb}
It holds  $\uparrow I^{-}(x)=\textrm{Int} \,D_p^{+}(x)$,
$\overline{D_p^{+}(x)}=A^{+}(x)$, and $\downarrow
I^{+}(x)=\textrm{Int} \,D_f^{-}(x)$,
$\overline{D_f^{-}(x)}=A^{-}(x)$.
\end{lemma}

\begin{proof}
It suffices to prove the characterization $D^{+}_p=\{(x,z) : \forall
s \in I^{-}(x), s \ll z \}$ which implies $D^{+}_p(x)=\{z \in M:
\forall s \in I^{-}(x), s \ll z \}$ and hence $\uparrow
I^{-}(x)=\textrm{Int} \,D_p^{+}(x)$. Indeed, if $(x,z) \in D^{+}_p$
then $x \in \overline{I^{-}(z)}$ thus taken $s \in I^{-}(x)$, $x \in
I^{+}(s)$ and since $I^{+}$ is open $s \ll z$, thus $D^{+}_p\subset
\{(x,z) : \forall s \in I^{-}(x), s \ll z \}$. Conversely, if $(x,z)
\in \{(x,z) : \forall s \in I^{-}(x), s \ll z \}$ then taken $s \in
I^{-}(x)$ it is $s \in I^{-}(z)$ and since $s$ can be chosen
arbitrarily close to $x$, $x \in \overline{I^{-}(z)}$, i.e. $\{(x,z)
: \forall s \in I^{-}(x), s \ll z \} \subset D^{+}_p$. The other
statements are proved analogously.
\end{proof}

%
%\begin{proof}
%It suffices to prove $\uparrow I^{-}(x)=\textrm{Int} \,D_p^{+}(x)$
%and $\downarrow I^{+}(x)=\textrm{Int} \,D_f^{-}(x)$. For the former
%case $\textrm{Int} \,D_p^{+}(x)=\textrm{Int}\{w \in M: x \in
%\overline{I^{-}(w)}\} $ thus any $y\in \textrm{Int} \,D_p^{+}(x)$
%has a neighborhood $W$ such that for every $w \in W$, $x \in
%\overline{I^{-}(w)}$, in particular if $z \in I^{-}(x)$, $w \in
%I^{+}(z)$ and thus $y \in \uparrow \! I^{-}(x)$, i.e. $ \textrm{Int}
%\,D_p^{+}(x) \subset \uparrow\! I^{-}(x)$. Conversely, if $y \in
%\uparrow I^{-}(x)$ there is a neighborhood $W\ni y$, such that for
%every $w \in W$ and  for every $z \in I^{-}(x)$, $w \in I^{+}(z)$
%i.e. $z \in I^{-}(w)$. Since $z$ is arbitrary, $x \in
%\overline{I^{-}(w)}$, hence $y \in \textrm{Int} D^{+}_p(x)$. A
%similar proof can be given for the $D_f^{-}(x)$ case taking into
%account that $\textrm{Int} \,D_f^{-}(x)=\textrm{Int}\{w \in M: x \in
%\overline{I^{+}(w)}\} $.
%\end{proof}

Although $\bar{J}^{+}$ is not necessarily transitive the following
result holds

\begin{theorem} \label{poa}
The causal relations $B_p^{+}=\{(x,y): y\in \,\uparrow\! I^{-}(x)\}$
and $B_f^{+}=\{(x,y): x\in\, \downarrow\! I^{+}(y)\}$ are
transitive, that is:
\begin{itemize}
\item[(a)] If $y \in\, \uparrow\! I^{-}(x)$ and $ z \in \,\uparrow\!
I^{-}(y) $ then $ z \in\, \uparrow \! I^{-}(x) $.
\item[(b)] If $x \in\, \downarrow \!I^{+}(y)$ and $ y \in\, \downarrow
\!I^{+}(z) $ then $ x \in\, \downarrow \!I^{+}(z) $.
\end{itemize}
Moreover, $\overline{B^{-}_p(y)}=\overline{I^{-}(y)}=D^{-}_p(y)$ and
$\overline{B^{+}_f(x)}=\overline{I^{+}(x)}=D^{+}_f(x)$.

\end{theorem}

\begin{proof}
Case $\uparrow \!I^{-}$, the other  case being analogous. Let $y
\in\, \uparrow\! I^{-}(x)$ and $ z \in \,\uparrow\! I^{-}(y) $.
Since $\uparrow\! I^{-}(x)=\textrm{Int}D_p^{+}(x)=\textrm{Int}\{w
\in M: x \in \overline{I^{-}(w)}\} $ there is a neighborhood $W\ni
y$ such that for all $w \in W$, $x\in \overline{I^{-}(w)}$.
Analogously there is a neighborhood $W'\ni z$ such that for every
$w' \in W'$, $y \in \overline{I^{-}(w')}$. Choose $w \in W \cap
I^{-}(y)$. Whatever the event $w' \in W'$,  we have the chain $x \in
\overline{I^{-}(w)}$, $(w,y) \in I^{+}$, $y\in
\overline{I^{-}(w')}$. Since $I^{+}$ is open $x \in
\overline{I^{-}(w')}$, and since $w' \in W'$ is arbitrary, and $W'$
is an open neighborhood of $z$, $z \in \uparrow\! I^{-}(x)$.

Let us come to the proof of
$\overline{B^{-}_p(y)}=\overline{I^{-}(y)}$. Let $x\in
\overline{B^{-}_p(y)}$, there are $x_n$, such that $x_n \to x$ and
$y \in \uparrow\! I^{-}(x_n) \subset D^{+}_p(x_n)$. Thus $x_n \in
\overline{I^{-}(y)}$ and hence $x \in  \overline{I^{-}(y)}$.
Conversely, if $x \in  \overline{I^{-}(y)}$ there are $x_n$, such
that $x_n \to x$ and $x_n \in I^{-}(y)$ or equivalently $y \in
I^{+}(x_n)\subset \uparrow\! I^{-}(x_n)$ which reads $(x_n,y)\in
B^{+}_p$ and finally $x \in \overline{B^{-}_p(y)}$. The proof of
$\overline{B^{+}_f(x)}=\overline{I^{+}(x)}$ is analogous.

\end{proof}

\begin{theorem}
If the spacetime $(M,g)$ is past distinguishing then $B_p^{+}$ is
antisymmetric. Analogously, if the spacetime $(M,g)$ is future
distinguishing then $B_f^{+}$ is antisymmetric.
\end{theorem}

\begin{proof}
It is a  consequence of the inclusions $\uparrow~\! I^{-}(x)\subset
\,D_p^{+}(x)$, and $\downarrow~I^{+}(x)\subset \,D_f^{-}(x)$, given
by lemma \ref{hyb}. Indeed, for instance, $(x,z)\in B_p^{+}$ and
$(z,x)\in B_p^{+}$ reads $ z\in \,\uparrow\! I^{-}(x)$ and  $ x\in
\,\uparrow\! I^{-}(z)$, hence $  z\in D_p^{+}(x)$ and $  x\in
D_p^{+}(z)$, which reads $(x,z) \in D_p^{+}$ and $(z,x) \in
D_p^{+}$, and using the antisymmetry of $D_p^{+}$, $x=z$.

\end{proof}

The point based relation $\uparrow \!I^{-}(x)$ (or $\downarrow
\!I^{+}(x)$) is also nicely related to strong causality. Indeed, I.
R\'acz has shown \cite[Prop. 3.1]{racz87} that the map $\uparrow
\!I^{-}: M \to P(M)$ (or $\downarrow \!I^{+}: M \to P(M)$) is
injective iff the spacetime is strongly causal.

Note that since $B^{+}_p$ is transitive $\uparrow\! I^{-}(x)$ is a
future set. Analogously, since $B^{+}_f$ is transitive,
$\downarrow\! I^{+}(x)$ is a past set. Thus using \cite[Prop.
3.7]{beem96}

\begin{align}
A^{+}(x)&=\overline{\uparrow\! I^{-}(x)}=\{y \in M: I^{+}(y)\subset
\uparrow\!I^{-}(x)\}, \\
A^{-}(x)&=\overline{\downarrow\! I^{+}(x)}=\{y \in M:
I^{-}(y)\subset \downarrow\! I^{+}(x)\},
\end{align}
which is the original, rather involved, definition of the sets
$A^{\pm}(x)$ given by Woodhouse \cite{woodhouse73}. Akolia et al.
\cite[Prop. 10]{akolia81}  noted, as also proved here, that the
causal relation so defined coincides with $\bar{J}^{+}$ which
explains why I started directly from this simpler definition of
$A^{+}$.

After the introduction of the properties of chronology and causality
by Kronheimer and Penrose \cite{kronheimer67} and strong causality
and stable causality by S. Hawking \cite{hawking68}, B. Carter
\cite{carter71} introduced the causal virtuosity hierarchy with the
aim of making some order in the different causality requirements
that were appearing in the literature. He showed that between strong
causality and stable causality  a denumerable sequence of, each time
more demanding, properties could be defined. Carter's definitions
were quite involved because he used the point based causal relation
$J^{\pm}(x)$ instead of the more versatile $J^{+}$.

He  defined a sequence of causal relations, let me denote them
${\le}^{n}$, in which $\le^0=\le$, and $\le^{n}$ was obtained by
taking suitable closures and compositions of the previous causal
relations $\le^{k}, k<n$  (for an account see
\cite{garciaparrado05}). According to Carter's definition a
spacetime is  {\em $n$-th degree causally virtuous} (sometimes
referred to as {\em $n$-th order strongly causal}), $n \ge 0$, if $x
\le^i z$ and $z \le^j x$ with $i+j=n$ implies $x=z$. In particular
it is {\em infinitely causally virtuous} if it is $n$-th degree
causally virtuous for every $n$, that is if, whatever $i,j \in
\mathbb{N}$, $x \le^i z$ and $z \le^j x$ implies $x=z$. According to
Carter, $0$-th degree causally virtuous spacetimes are simply the
causal spacetimes, first causally virtuous spacetimes are the
distinguishing spacetimes and the second causally virtuous
spacetimes are the strongly causal spacetimes. That $n$-th degree
causal virtuosity is different from $n+1$-th degree causal
virtuosity was shown in an example due to Carter and published in
\cite[Fig. 25]{penrose72}.

It must be said that given the causal hierarchy there is essentially
no proof that the hierarchy is complete and a statement of this kind
would probably make no sense at all. It can always happen that some
day a new interesting causal property could be discovered which fits
nicely in the hierarchy and simplifies some old statements and
proofs. Carter's causal ladder had the merit to clarify this point
but, at least in the author's opinion, the new levels introduced by
Carter failed to prove particularly useful for the development of
causality theory. In this respect $K$-causality is conceptually
simpler but very similar to the {\em infinitely causally virtuous}
property (if they are equivalent a proof would probably be
complicated by the involved definition of the latter). It conveys
the same ideas in a simplified way, and I think an almost definitive
causal ladder should accommodate it  in place of the causal
virtuosity (sub)ladder.

The analogy between infinite causal virtuosity and $K$-causality
becomes even more stringent if one recalls that the set $K^{+}$ can
be built starting from $J^{+}$ via a {\em transfinite} induction
\cite[Lemma 14]{sorkin96} in which at each step new pairs of events
in the closure or obtained through transitivity are added, in a way
which clearly resembles that used by Carter for the definition of
his $\le ^n$ relations, but with the advantage that here no point
based causal relation is used.

It is easy to prove the following

\begin{theorem} \label{nsx}
$K$-causality implies infinite causal virtuosity.
\end{theorem}

\begin{proof}
The starting point of Carter's inductive process, i.e. $J^{+}$, is
contained in $K^{+}$, and the construction of the sets corresponding
to $\le^k$, $k>0$, is obtained through compositions and closures
that, due to the transitivity and closure properties of $K^{+}$
necessarily remain included in $K^{+}$. Thus $x \le^n z \Rightarrow
(x,z) \in K^{+}$ and hence $K$-causality implies infinite causal
virtuosity.
\end{proof}

The short account given by Penrose \cite[Remark 4.19]{penrose72} of
Carter's causal ladder has introduced some terminological confusion.
He recognized that the main point of Carter's analysis was the
possibility of constructing an infinite causal ladder between strong
causality and stable causality, however instead of working with
Carter's involved definitions he considered a simplified causal
ladder which actually did not coincide with Carter's as it had wider
steps (for instance the distinguishing property was not included).
Unfortunately, due to this account, sometimes Penrose's causal
ladder is identified with that introduced by Carter (see, for
instance, \cite{racz87}) a fact which may arise some confusion.

Penrose's ideas anticipated those by Woodhouse. Essentially, he
considered  a generalization of the notion of $A$-causal spacetime
to arbitrary chains. In order to keep the connection with the
$A$-causality property and the causal relations on $M \times M$ it
is convenient to introduce Penrose's ladder as follows

\begin{definition} \label{mjx}
The set $A^{+ n}\subset M\times M$, $n\ge 1$, is the set of pairs
$(x_1,x_{n+1})$, which can be connected by a $n$-chain
$(x_i,x_{i+1}) \in A^{+}$, $i=1 \ldots n$, and $A^{+0}=\Delta$. In
particular $A^{+1}=A^{+}$. The set $A^{+\infty}=\cup^{+\infty}_{i=0}
A^{+i}$,  is the set of the pairs of events which are connected by a
chain of $A$-causally related events.

A spacetime $(M,g)$ is {\em $A^{n/2}$-causal}, $n\ge 2$, if the
existence of a cyclic $n$-chain $(x_i,x_{i+1}) \in A^{+}$, $i=1
\ldots n$, $x_1=x_{n+1}$, implies $x_1=x_i$, $i=1, \ldots, n$. A
spacetime is {\em $A^{\infty}$-causal} if it is $A^{k/2}$-causal for
every integer $k\ge 2$ (or, equivalently, if $A^{+\infty}$ is
antisymmetric).
\end{definition}

This definition is motivated by the fact that if $n$ is even then
the $A^{n/2}$-causality property coincides\footnote{This statement
can be proved as follows. Assume $n$ even, and let the spacetime be
$A^{n/2}$-causal according to definition \ref{mjx}, let $(x,z) \in
A^{+n/2}$ and $(z,x) \in A^{+n/2}$ then there is a $n$-chain of
$A^{+}$-related events connecting $x$ to itself passing through $z$,
thus $x=z$. Conversely, if the spacetime is $A^{n/2}$-causal
according to section \ref{mja}, then given  a cyclic $n$-chain
$(x_{i},x_{i+1})$, $x_{n+1}=x_1$, which connects $x_1$ to itself
then $(x_1, x_{n/2+1}) \in A^{+n/2}$ and $(x_{n/2+1}, x_{n+1}) \in
A^{+n/2}$ which implies $x_{n/2+1}=x_1$. Now use the fact that
$A^{+}$ is reflexive, and repeat the argument to obtain that
$x_i=x_1$.} with the requirement of antisymmetry for the $A^{+n/2}$
causal relation in agreement with the general definitions of section
\ref{mja}. If instead $n$ is odd there is no clear correspondence
with a set on $M\times M$, and indeed no set $A^{+n/2}\subset
M\times M$ has been defined for odd $n$. Clearly, since $A^{+}$ is
reflexive $A^{(k+1)/2}$-causality implies $A^{k/2}$-causality,
moreover they are distinct properties due to the usual example
\cite[Fig. 25]{penrose72}. Note also that $A^{+\infty}$ is
transitive but not closed.

Since $A^{+} \subset K^{+}$, it is $A^{+n}\subset K^{+}$ and hence

\begin{theorem}
$K$-causality implies $A^{\infty}$-causality.
\end{theorem}

After the introduction of the infinitely causally virtuous property,
S. Hawking  \cite{hawking71} considered the possibility of its
coincidence with the previously defined stably causal property. He
expressed the opinion that this coincidence does not hold,
 without, as far as I know, providing an example of spacetime
infinitely causally virtuous but non-stably causal. As I said any
statement regarding Carter's causal properties is in general
difficult to prove because of their involved definitions. It is then
better to work with the $A^{k/2}$-causal ladder. In this respect it
is meaningful to ask whether $A^{\infty}$-causality coincides with
$K$-causality. The spacetime example I provide in section \ref{bap}
is $A^{\infty}$-causal but is not $K$-causal.

\section{A spacetime example} \label{bap}
\begin{figure}
\centering \psfrag{x}{$x$} \psfrag{z}{$\!z$} \psfrag{g}{$\!\!
\gamma$} \psfrag{a}{$a$} \psfrag{b}{$b$} \psfrag{c}{$c$}
\psfrag{xn}{$x_n$} \psfrag{zn}{$z_n$} \psfrag{yn}{$y_n$}
\includegraphics[width=13cm]{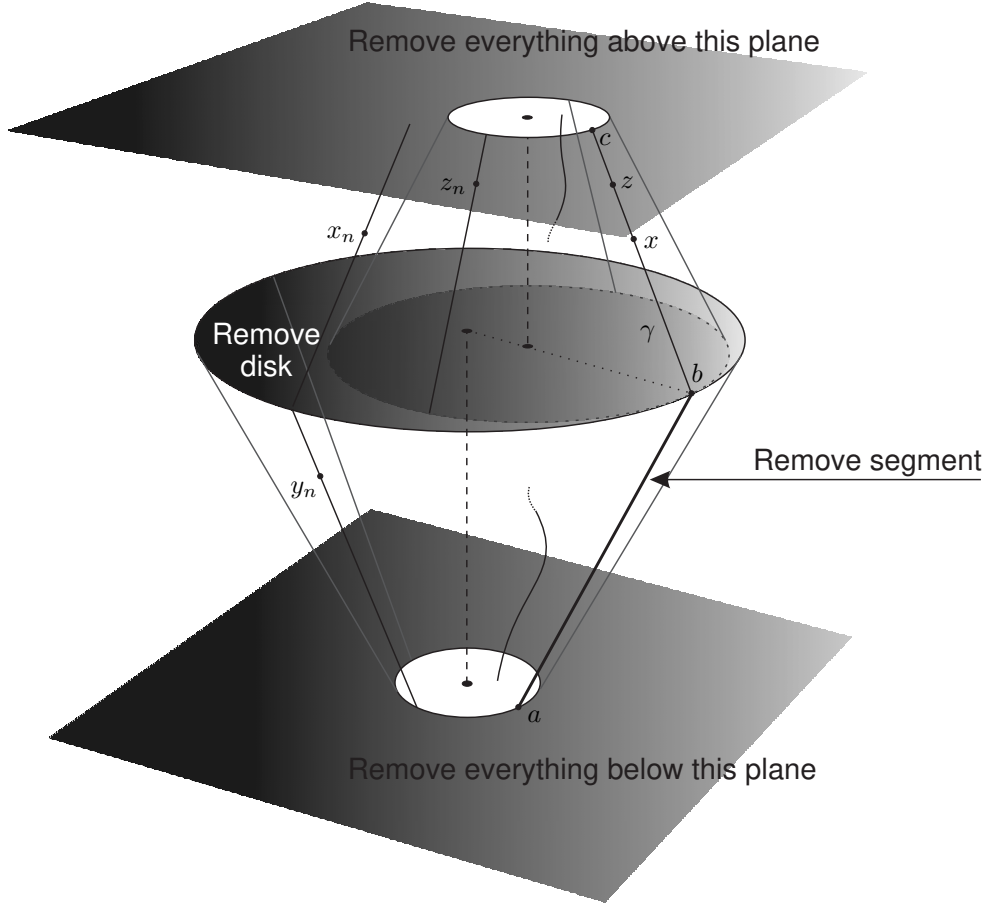}
\caption{A non-$K$-causal but $A^{\infty}$-causal spacetime is
obtained from Minkowski spacetime by removing a disk, a geodesic
segment, two planes with holes of the same size and by identifying
the holes as shown in the figure. The metric is that induced from
Minkowski spacetime. The events $x_n$, $y_n$ and $z_n$ are chosen so
that $(z_n,y_n) \in \bar{J}^{+}$, thanks to the identification of
the holes, and $(y_n,x_n) \in \bar{J}^{+}$. Thus $(z_n,x_n) \in
K^{+}$ although $(x_n,z_n) \notin \bar{J}^{+}$. Since $(z,x)$ can be
regarded as a limit of a suitable sequence $(z_n,x_ n)$, it is
$(z,x) \in K^{+}$ and also $(x,z) \in J^{+}$. } \label{fig1}
\end{figure}

In this section I give an example of $A^{\infty}$-causal
non-$K$-causal
 spacetime (see figure \ref{fig1}).

Let $(\Lambda,\eta)$ be 2+1 Minkowski spacetime and let $(t,x^1,
x^2)$ be canonical coordinates, $\eta=-\dd t^2+(\dd x^{1})^2 +(\dd
x^{2})^2$. The manifold $M$ is obtained from $\Lambda$ as follows.
Remove the planes $t=0$ and,  $t=1$. On the planes there are two
(open in the plane topology) holes of the same size but with
non-aligned centers which are not removed but rather identified (the
correspondence between points is done respecting the Minkowskian
parallel transport). The causal future of the lower hole is
`stopped' by a (closed) disk removed from Minkowski spacetime and of
exactly the same size of the light cone at that height. The height
of the disk from $t=0$ is chosen so that the causal past of the
upper hole reaches the edge of the removed disk at a point $b$ (also
removed). The lightlike geodesic segment $ab$ in the boundary of the
light cone issuing from the lower hole is also removed. The metric
$g=\eta \vert_{M}$ is that induced from Minkowski spacetime.

A close inspection of this spacetime shows that if it were
non-$A^{\infty}$-causal then the two events $x,z \in M$ such that
$(x,z) \in A^{+\infty}$, $(z,x) \in A^{+\infty}$ would necessarily
stay in the lightlike inextendible geodesic $\gamma$ obtained from
the segment $bc$ after removal of the endpoints.  We may assume $z
\in J^{+}(x)$. However, it is not possible that $x \in
A^{+\infty}(z)$ indeed the set $A^{-}(x)=\{y: (y,x) \in
\bar{J}^{+}\} $ is closed but has empty intersection with the
closure of the causal future of the hole, thanks to the fact that
the segment $ab$ has been removed. Thus no matter how long is the
chain of $A$-causally related events considered, none can connect
$z$ to $x$, thus $(z,x) \notin A^{+\infty}$ although $(z,x) \in
K^{+}$ as the argument of the figure caption shows. Thus the
spacetime is $A^{\infty}$-causal and non-$K$-causal.

\section{Conclusions}

In this work a unifying approach to the study of causal relations
has been presented in which the associated antisymmetry conditions
play an important role. Indeed, through them  the construction of
the causal ladder becomes particularly clear as the relationship
between the different causality requirements follows trivially from
the inclusion of sets in $M\times M$.

Some new results on causality theory have been obtained. Among them
the equivalence between openness (closure) and partial openness
(partial closure) of transitive causal relations. The equivalence
between antisymmetry and injectivity for reflexive and transitive
relations. Some new characterization of $K$-causality and the
$K^{+}$ relation, i.e. corollaries \ref{co1} and \ref{co2}, or
strong causality (theorem \ref{cvd}). The fact that the past and
future distinction properties can be characterized through the
antisymmetry conditions of suitable transitive and reflexive causal
relations $D^{+}_p$ and $D^{+}_f$ (theorem \ref{mka}).

Finally, other causal relations have been studied pointing out
whether they are transitive or not, closed or open, partially closed
or partially open. The $A$-causality subladder has been presented in
detail and in the last section a spacetime example has been given
which proves that $K$-causality differs from infinite $A$-causality.

\section*{Acknowledgements}
I warmly thank F. Dowker, R. Low, R.D. Sorkin and E. Woolgar, for
comments on the spacetime example of section \ref{bap}   and D.
Canarutto for help in  drawing the figure. Finally, I thank I.
R\'acz for pointing out reference \cite{racz87}.

%\bibliography{../../bibliografie/simultaneity,../../bibliografie/libri,../../bibliografie/miei,../../bibliografie/mieiPreprints,../../bibliografie/mieiProceedings}

\begin{thebibliography}{10}

\bibitem{akolia81}
G.~M. Akolia, P.S. Joshi, and U.D. Vyas.
\newblock On almost causality.
\newblock {\em J. Math. Phys.}, 22:1243--1247, 1981.

\bibitem{beem96}
J.~K. Beem, P.~E. Ehrlich, and K.~L. Easley.
\newblock {\em Global Lorentzian Geometry}.
\newblock Marcel {D}ekker {I}nc., New York, 1996.

\bibitem{carter71}
B.~Carter.
\newblock Causal structure in space-time.
\newblock {\em Gen. Relativ. Gravit.}, 1:349--391, 1971.

\bibitem{dowker00}
H.~F. Dowker, R.~S. Garcia, and S.~Surya.
\newblock {$K$}-causality and degenerate spacetimes.
\newblock {\em Class. Quantum Grav.}, 17:4377--4396, 2000.

\bibitem{garciaparrado05}
A.~Garc{\'\i}a-Parrado and J.~M.~M. Senovilla.
\newblock Causal structures and causal boundaries.
\newblock {\em Class. Quantum Grav.}, 22:R1--R84, 2005.

\bibitem{hawking68}
S.~W. Hawking.
\newblock The existence of cosmic time functions.
\newblock {\em Proc. {R}oy. {S}oc. {L}ondon, series {A}}, 308:433--435, 1968.

\bibitem{hawking71}
S.~W. Hawking.
\newblock Stable and generic properties in general relativity.
\newblock {\em Gen. Relativ. Gravit.}, 1:393--400, 1971.

\bibitem{hawking74}
S.~W. Hawking and R.~K. Sachs.
\newblock Causally continuous spacetimes.
\newblock {\em Commun. Math. Phys.}, 35:287--296, 1974.

\bibitem{hrbacek99}
K.~Hrbacek and T.~Jech.
\newblock {\em Introduction to set theory}.
\newblock {M}arcel {D}ekker, New York, 1999.

\bibitem{kronheimer67}
E.~H. Kronheimer and R.~Penrose.
\newblock On the structure of causal spaces.
\newblock {\em Proc. {C}amb. {P}hil. {S}oc.}, 63:482--501, 1967.

\bibitem{minguzzi07}
E.~Minguzzi.
\newblock The causal ladder and the strength of  {$K$-causality}. II
\newblock {\em Class. Quantum Grav.} 2007, In press.


\bibitem{minguzzi06c}
E.~Minguzzi and M.~S\'anchez.
\newblock {\em The causal hierarchy of spacetimes}.
\newblock Cont. to Proc. of the ESI Semester `Geometry of pseudo-Riemannian
  Manifolds with Application to Physics', ed. D. Alekseevsky and H. Baum (ESI,
  Vienna, Sept–-Dec 2005) (European Mathematical Society Publishing House),
  gr-qc/0609119, To appear.

\bibitem{oneill83}
B.~{O'N}eill.
\newblock {\em Semi-{R}iemannian Geometry}.
\newblock Academic {P}ress, San Diego, 1983.

\bibitem{park97}
J.~C. Park.
\newblock Almost causal structure in spacetimes.
\newblock {\em J. {K}orean {M}ath. {S}oc.}, 34:257--264, 1997.

\bibitem{penrose72}
R.~Penrose.
\newblock {\em Techniques of Differential Topology in Relativity}.
\newblock Cbms-{N}sf Regional Conference Series in Applied Mathematics. {SIAM},
  Philadelphia, 1972.

\bibitem{racz87}
I.~R{\'a}cz.
\newblock Distinguishing properties of causality conditions.
\newblock {\em Gen. Relativ. Gravit.}, 19:1025--1031, 1987.

\bibitem{sorkin96}
R.~D. Sorkin and E.~Woolgar.
\newblock A causal order for spacetimes with {$C^0$} {L}orentzian metrics:
  proof of compactness of the space of causal curves.
\newblock {\em Class. Quantum Grav.}, 13:1971--1993, 1996.

\bibitem{woodhouse73}
N.~M.~J. Woodhouse.
\newblock The differentiable and causal structures of space-time.
\newblock {\em J. Math. Phys.}, 14:495--501, 1973.

\end{thebibliography}
%\bibliographystyle{plain}

\end{document}